\documentclass[conference]{IEEEtran}
\IEEEoverridecommandlockouts
\usepackage{cite}
\usepackage{amsmath,amssymb,amsfonts}
\usepackage{algorithmic}
\usepackage{graphicx}
\usepackage{textcomp}
\usepackage{xcolor}
\def\BibTeX{{\rm B\kern-.05em{\sc i\kern-.025em b}\kern-.08em
    T\kern-.1667em\lower.7ex\hbox{E}\kern-.125emX}}
\begin{document}

\title{Finder: A Multimodal AI-Powered Search Framework for Pharmaceutical Data Retrieval  \\
}

\author{
\IEEEauthorblockN{Suyash Mishra}
\IEEEauthorblockA{
Researcher, Global Product Strategy\\
F. Hoffmann-La Roche Ltd.\\
Basel, Switzerland\\
suyash.mishra@roche.com}
\and
\IEEEauthorblockN{Srikanth Patil}
\IEEEauthorblockA{
Associate Vice President (Gen AI)\\
Involead Services Pvt Ltd.\\
Pune, India\\
srikanth.patil@involead.com}
\and
\IEEEauthorblockN{Satyanarayan Pati}
\IEEEauthorblockA{
Lead Data Scientist\\
Involead Services Pvt Ltd.\\
Delhi, India\\
satyanarayan.pati@involead.com}
\and
\IEEEauthorblockN{Sagar Sahu}
\IEEEauthorblockA{
Data Scientist\\
Involead Services Pvt Ltd.\\
Bhubaneswar, India\\
sagar.sahu@involead.com}
\and
\IEEEauthorblockN{Baddu Narendra}
\IEEEauthorblockA{
Data Scientist\\
Involead Services Pvt Ltd.\\
Bhubaneswar, India\\
baddu.narendra@involead.com}

}

% Note: adjust 0.7\linewidth if needed to visually align the second-row first author
% with Suyash Mishra on the first row, depending on your final column width.

\maketitle

\begin{abstract}
AI is transforming pharmaceutical search, where traditional systems struggle with multimodal content and manual curation. Finder is a scalable AI-powered framework that unifies retrieval across text, images, audio, and video using hybrid vector search, combining sparse lexical and dense semantic models.
Its modular pipeline ingests diverse formats, enriches metadata, and stores content in a vector-native backend. Finder supports reasoning-aware natural language search, improving precision and contextual relevance.
The system has processed over 291,400 documents, 31,070 videos, and 1,192 audio files in 98 languages. Techniques like hybrid fusion, chunking, and metadata-aware routing enable intelligent access across regulatory, research, and commercial domains.
\end{abstract}

\section{Introduction}

Pharmaceutical companies are overwhelmed by the exponential increase in data, the diversity of modalities, and the complexity of regulations, which are the results of advances in biomedicine and the digitization of compliance workflows. The need for rapid access to unstructured content, clinical reports, filings, training materials, and recordings is the key to agility and innovation ~\cite{b1}.

If we look at traditional keyword-based search systems we can see that they cannot handle multimodal content, which means that they are not able to support some formats like annotated slides, transcribed audio, and visual data. The outcome of it is fragmented access, low quality of metadata, and the necessity of a lot of manual work in the processes such as pharmacovigilance and audits ~\cite{b2}.

AI advancements have greatly improved the capabilities of multimodal understanding, and it can now handle natural language reasoning as well~\cite{b3}. Such new abilities help a lot when trying to find information in various formats like graphs, scanned documents, and video annotations.

 This paper introduces \textit{Finder}, an AI-powered multimodal
search framework tailored for pharmaceutical enterprises.
Finder ingests content across formats (PDF, Word, Power-
Point, audio, video, and images) and converts it into struc-
tured textual representations using a modular pipeline. Com-
ponents such as OpenAI Whisper for audio transcription
and Qwen2 for video captioning are integrated to support
comprehensive content normalization. The processed data
is enriched with metadata and stored in a document-aware
backend, enabling hybrid search capabilities that combine
sparse lexical models (BM42) with dense semantic embed-
dings (Mixedbread). These models are fused using reciprocal
rank optimization to enhance retrieval performance.

Compared to task-specific solutions that are only focused on chatbot assistance~\cite{b4}, audio summarization~\cite{b5}, or visual parsing~\cite{b6}, Finder is a single search layer that comprises of scoring for relevance, document chunking, and modality-aware query interpretation. It allows exact retrieval without the need of keyword expertise and is developed to be combined with entity recognition for better context.

Finder makes content management better in pharmaceutical environments through metadata-aware routing, user intent inference, and natural language optimization. It allows the accurate retrieval of clauses, versioned documents, and annotated visuals without the need of explicit keywords.

Deployed across enterprise environments, Finder supports use cases like audits, literature mining, training retrieval, and cross-functional discovery. Its scalable architecture enables retrieval-augmented generation (RAG), positioning it for generative search and decision-support in healthcare.

This paper:
\begin{itemize}
    \item Presents Finder’s architecture and ingestion pipeline
    \item Describes its hybrid search methodology
    \item Evaluates retrieval performance
    \item Outlines future enhancements including re-ranking and conversational retrieval
\end{itemize}

Finder demonstrates how AI-powered search can unlock organizational knowledge, reduce discovery time, and improve decision-making in regulated domains.

\section{Related Works}

\subsection{Vision-Language and Multimodal Models}

Vision-language models (VLMs) jointly process images and text for tasks such as image captioning and visual question answering. The most famous models are CLIP (OpenAI)~\cite{b7}, BLIP (Salesforce)~\cite{b8}, and Flamingo (DeepMind)~\cite{b9}. GPT-4V~\cite{b10} and Google Gemini~\cite{b11} are examples of such systems that recently have been developed to extend these faculties to chatbots, so that by a combined image and a language encoder they could receive a visual query and handle it.

\subsection{Lexical vs. Neural Retrieval}

Lexical methods like BM25~\cite{b12} rank documents on the basis of term frequency–inverse document frequency, which are good with exact matches. Neural retrievers (e.g., BERT, DPR~\cite{b13}) create dense embeddings that are used for semantic search. Models like SPLADE~\cite{b14} merge sparse lexical signals with learned semantic features trying to get higher recall.

\subsection{Hybrid Retrieval}

Hybrid retrieval combines the advantages of both lexical and semantic to improve performance. Score fusion (BM25 + embedding similarity\cite{b15}), hybrid indexing (dense-sparse vectors via ANN), and query expansion (e.g., Multi-Query, HyDE~\cite{b16}) are three different techniques that could be used to make retrieval better. The trend of domain-specific fine-tuning is still more than prevalent for technical fields.

\subsection{LLM-based and Domain-Specific Search} 

RAG (Retrieval-Augmented Generation) links the outputs of LLM to external sources. QA-RAG~\cite{b17} in medical field takes user queries and LLM-generate answers, then searches for regulation documents with matching content. Baidu’s EICopilot~\cite{b18}, on the other hand, converts natural queries into graph database queries (Gremlin~\cite{b19}) making structured data retrieval more efficient.

\subsection{Search Systems and Infrastructure}

The current search stacks are utilizing vector databases and approximate nearest neighbor (ANN) algorithms. One of the main drivers of the dense retrieval ecosystems like Pinecone and Zilliz is the FAISS~\cite{b20} which allows efficient GPU/CPU indexing for dense retrieval. Various methods such as HNSW and IVFPQ have made it possible to conduct similarity search on large volumes of data. The use of hybrid search modes and benchmarks (e.g., MS MARCO, Natural Questions) not only assists in measuring the effectiveness but also helps in determining the performance of BM25 and neural rankers.

\section{Methodology}

This section briefly describes the architecture, intelligent pipelines, and AI-integrated modules that are part of the  \textit{Finder} platform. Enterprise-scale multimodal retrieval capabilities such as ingestion, normalization, vectorization, hybrid search, and real-time ranking over large pharmaceutical corpora are all supported by Finder. 

\begin{figure}[htbp]
\centering
\includegraphics[width=\linewidth]{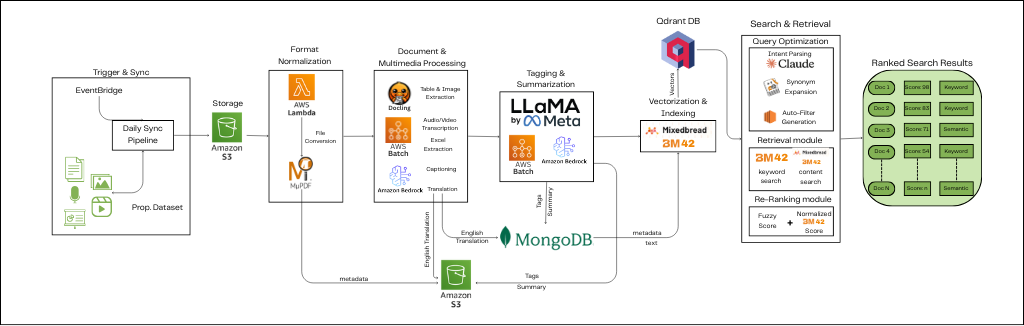}  % make sure the file is uploaded with this exact name
\caption{Finder Architecture Diagram}
\label{fig:finder-architecture}  % descriptive label for referencing
\end{figure}

\subsection{System Overview}

Finder is a modular hybrid search engine that utilizes a combination of lexical and semantic techniques to retrieve information from knowledge-intensive domains. It can take in data in any of the mentioned formats, that is, PDF, PowerPoint, Excel, images, audio, and video, then normalize it, enrich with semantic tags, and store textual as well as vector forms in a scalable backend. Retrieval proceeds by tuning queries with a large language model (Claude 4), automatically inferring the filters, and running a hybrid lexical-semantic search with BM42 and Mixedbread. Weighted fusion of fuzzy matching and normalized BM42 scores is used in the final re-ranking stage to emphasize the contextually relevant results. 

\subsection{Data Ingestion and Format Normalization}

Documents are uploaded to Amazon S3 using a standardized object key schema:
\begin{itemize}
    \item \texttt{s3://bucket/raw/\{filename\}} for unprocessed inputs
    \item \texttt{s3://bucket/processed/\{filename\}} for normalized outputs
\end{itemize}

Files that are not in PDF format (for example, DOCX, PPTX) are converted to PDF with the help of PyMuPDF. Metadata such as the author, date of creation, and annotations are also transferred. Feature extraction is done with Docling, a tool created by IBM Research, that collects document-level features for the following stages of filtering and semantic analysis. 

\subsection{Document and Multimedia Processing}

\begin{itemize}
    \item \textbf{Text, Table, and Image Extraction:} Docling is utilized to parse each page of the document with the aim of separately extracting the text, tables, and images. These extracted parts are saved in S3 for modular access.
    \item \textbf{Image Captioning:} Pictures are treated with AWS Bedrock (\texttt{amazon.nova-lite-v1:0})~\cite{b24} to get simplified captions that merge the visual and the textual context.
    \item \textbf{Language Detection and Translation:} Langdetect is used to identify non-English inputs and AWS Bedrock is used for translation.
\end{itemize}

\subsection{Tagging Pipeline}

\begin{itemize}
    \item \textbf{Extractive Tagging:} A knowledge base with domain-specific content is utilized to fetch the tags from the pharmaceutical area, for instance, product, indication, and topic.
    \item \textbf{Abstractive Tagging:} Meta-LLaMA 3 prompts generate high-level labels like content purpose and target audience, each with a confidence score.
    \item \textbf{Document Summarization:} Latencies to document summaries are made through Meta-LLaMA 3 via AWS Bedrock.
    \item \textbf{Tag Storage:} Tags along with their provenance are stored in JSON format to enrich metadata and improve search relevance.
\end{itemize}

\subsection{Vectorization}

\begin{figure}[htbp]
\centerline{\includegraphics[width=1\linewidth]{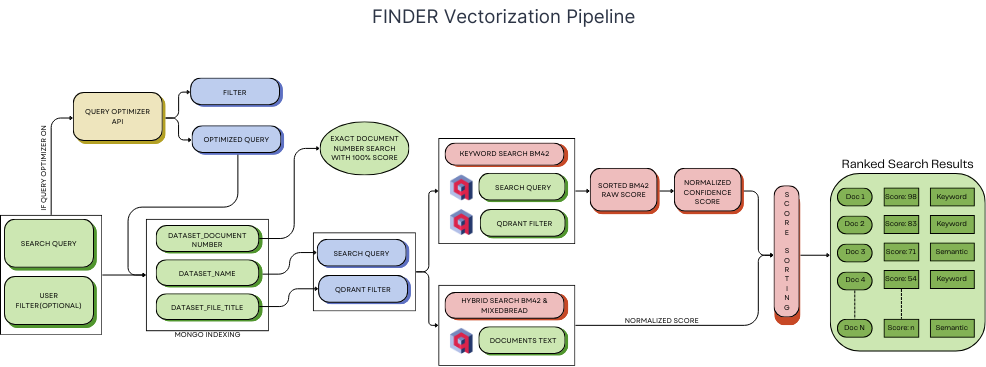}}
\caption{Vectorization Pipeline}
\label{fig}
\end{figure}

Finder’s vectorization pipeline fuses lexical and semantic searches to achieve high precision in retrieval. It starts with a user query which can be optionally filtered by product, region, or language. The Query Optimizer driven by Claude 4, if activated, reforms the query by way of decomposition, rewriting, and metadata enrichment.

The composite query is indexed in MongoDB using metadata fields such as \texttt{dataset\_document\_number}, \texttt{dataset\_name}, and \texttt{dataset\_file\_title}. The system thus, in the case of finding an exact match, assigns a 100\% relevance score and further search is skipped. If no exact match is found, the query is passed through Qdrant’s filter engine for vector-based retrieval.

The keyword search is conducted via BM42 over metadata fields. The search outputs raw scores that are then normalized into confidence values. At the same time, the hybrid search makes use of Mixedbread embeddings that merge with BM42 scores on document text. All the results are normalized and ranked, thus, a final sorted list based on relevance and modality is obtained.

\subsection{AI-Powered Search Workflow}

\begin{itemize}
    \item \textbf{Intent Parsing:} Claude 4 understands the user queries by converting them into a structured JSON format which includes \texttt{main\_query}, \texttt{filters}, and reformulations.
    \item \textbf{Glossary and Synonym Expansion:} A dynamic glossary changes medical abbreviations (e.g., ``IBD'' $\rightarrow$ ``Inflammatory Bowel Disease'') by using cosine similarity over embedded vectors.
    \item \textbf{Auto-Filter Module:} In case filters are absent, required fields such as country, language, product name, scientific area, and content type are automatically guessed. The filters set by the user take precedence over those generated by the system.
\end{itemize}

\subsection{Ranking Strategy: Hybrid Retrieval and Scoring Fusion}

Finder ranks using methods that are the opposite of one another, as it takes into account both keywords based on lexical matching and also semantic vector search. BM42 calculates lexical scores over structured metadata fields, whereas Mixed Bread embeddings provide contextual relevance from the document text.

Each document’s BM42 score is normalized using:
\begin{equation}
\text{Normalized Score} = \frac{\text{Raw Score}}{\text{Max Score}}
\label{eq:score}
\end{equation}

In order to increase relevancy, Finder employs fuzzy matching between the query and the document title using the \texttt{token\_set\_ratio} function from the RapidFuzz library~\cite{b25}. The final ranking score is calculated by using a weighted fusion: 30\% is assigned to the fuzzy score and 70\% to the normalized BM42 score. At the moment, there is a learning-to-rank (LTR) model that is being developed to optimize this fusion dynamically based on user feedback and behavioral signals.

\subsection{Monitoring and Evaluation Framework}

\begin{itemize}
    \item \textbf{Offline Benchmarking:} The evaluation of the performance is done based on a dataset of user-tests specific for the pharmaceutical domain. The metrics used are nDCG@10, Recall@50, and MRR.
    \item \textbf{Performance SLAs:} The monitoring of these is accomplished by the SLI/SLO pipelining that keeps track of end-to-end query latency, ingestion time, as well as the success/failure rates.
    \item \textbf{Failure Analysis:} In this case, the sources of failure like empty results, ASR misfires, and parsing issues have to be logged and further analyzed to be us
\end{itemize}

\section{Experimental Evaluation and Observations}

This section outlines the data employed by the \textit{Finder} platform to test its performance in enterprise-scale pharmaceutical environments, along with the evaluation setup, and tools.

\subsection{Proprietary Pharmaceutical Dataset and Scale}

Finder was evaluated on a proprietary, industry-regulated dataset spanning five years (July 2019–August 2025) and covering over 14 disease areas. The dataset includes:

\begin{itemize}
    \item \textbf{Documents:} Regulatory, safety, and scientific reports
    \item \textbf{Slides:} Marketing and training presentations
    \item \textbf{Images:} Scanned forms, diagrams, and annotated visuals
    \item \textbf{Audio:} Multilingual dictations and interviews
    \item \textbf{Video:} Educational content, key opinion leader (KOL) talks, and compliance briefings
\end{itemize}

Each file is tagged with metadata fields such as \texttt{document\_type}, \texttt{product}, \texttt{scientific\_area}, \texttt{language}, \texttt{region}, and \texttt{content\_purpose}. \textit{Note: The dataset is confidential and not publicly accessible.}

\textbf{Scale Overview:}
\begin{itemize}
    \item Total Documents: 314,343+
    \item Slide Decks: 78,028
    \item Videos: 31,070 (across 8+ formats)
    \item Audio Files: 1,192 (in 98+ languages)
    \item Images: 34,797
\end{itemize}

\textbf{Filetype Breakdown:}
\begin{itemize}
    \item PDF: 249,368
    \item PPT: 78,028
    \item Word: 40,706
    \item Text: 1,334
    \item Email: 193
    \item Excel: 828
    \item Others: 21,914
\end{itemize}

\subsection{Models and Tools Used}

\begin{table}[htbp]
\caption{Models and Tools Used in Finder}
\label{tab:models_tools}
\begin{center}
\begin{tabular}{|p{2.25cm}|p{2.25cm}|p{2.5cm}|}
\hline
\textbf{Component} & \textbf{Model/Tool} & \textbf{Notes} \\
\hline
Speech-to-Text & OpenAI Whisper & Automatic speech recognition (ASR) across 98 languages. \\
\hline
Extraction & PyMuPDF, Docling & Extracts text, tables, and images from each page for modular processing. \\
\hline
Tagging & AWS Bedrock: meta.llama3-8b-instruct-v1:0 & Assigns metadata labels with confidence scores. \\
\hline
Captioning & AWS Bedrock: amazon.nova-lite-v1:0 & Generates visual-textual summaries from images and tables. \\
\hline
Translation & AWS Bedrock: amazon.nova-micro-v1:0 & Enables multilingual document support. \\
\hline
Video Understanding & Qwen/Qwen2-VL-7B-Instruct & Performs frame-level action and scene summarization. \\
\hline
Semantic Embeddings & Mixedbread / SBERT & Used for dense vector search. \\
\hline
Sparse Retrieval & BM42 & Lexical keyword-based retrieval. \\
\hline
Query Optimization & Claude 4 & LLM-based intent parsing and auto-filter generation. \\
\hline
Vector Database & Qdrant & Supports real-time hybrid retrieval across sparse and dense vectors. \\
\hline
\end{tabular}
\end{center}
\end{table}

\subsection{Limitations and Observations}

In spite of several limitations being observed during the deployment phase, the performance was still solid:

\begin{itemize}
    \item \textbf{Translation Coverage:} The LLM-based translation pipeline might not fully cover certain low-resource languages. As a result, there may be multilingual coverage areas where the system can temporarily be silent. These translation gaps are largely overcome by rolling out manual quality assurance checks from time to time.
    \item \textbf{Video Scalability:} The latency overhead caused by scene understanding in long videos (exceeding 90 minutes) is what is limiting real-time responsiveness.
    \item \textbf{Inference Latency:} Transformer-based query expansion that normally takes milliseconds of additional processing time for long or complex queries is the reason. They are working on implementing caching and early pruning strategies to lessen this detrimental effect.
\end{itemize}

Next changes will be about fine-tuning the model for captioning and entity-linking to practically facilitate semantic accuracy and cut down on the waiting time.

\subsection{Retrieval Engine Evolution}

The Finder retrieval engine went through several redesigns which made not only the precision but also the ranking consistency better gradually:

\begin{table}[htbp]
\caption{Retrieval Engine Evolution}
\label{tab:retrieval_evolution}
\begin{center}
\begin{tabular}{|p{1.5cm}|p{1.5cm}|p{1.5cm}|p{2.5cm}|}
\hline
\textbf{Version} & \textbf{Keyword Search} & \textbf{Semantic Search} & \textbf{Remarks} \\
\hline
Initial & TF-IDF (MongoDB) & Mixedbread Large~\cite{b26} & Baseline performance with limited semantic coverage. \\
\hline
Intermediate & BM25~\cite{b27} & Mixedbread Large~\cite{b26} & Improved contextual matching and recall. \\
\hline
Current & BM42~\cite{b28} & BM42 + Mixedbread & Highest observed precision and lowest latency. \\
\hline
\end{tabular}
\end{center}
\end{table}

The first changes to be realized are the domain-specific fine-tuning for image captioning, tag generation, and entity linking. To diminish the semantic ambiguity and raise retrieval accuracy, subsequent releases will mount investigations into multi-vector embedding models like ColBERT and ColPaLI. Moreover, the next stage in the research will be the MuVERA integration to shorten latency and speed up results retrieval.

\subsection{Semantic Certainty Benchmarking}

Standard metrics such as MAP@5, MRR@5, and nDCG@5 are commonly used in retrieval evaluation and allow overall performance to be assessed. However, these metrics sometimes mask differences in performance for different query types. A new study~\cite{b29} reveals that retrieval correctness is linked to the query’s semantic certainty, where the latter is characterized by the geometrical stability and the neighborhood density of the embedding of the query vector in the vector space.
Using this framework, we benchmarked Finder across three pharmaceutical enterprise query types:

\begin{itemize}
    \item \textbf{Factual Queries} (e.g., "FDA-approved dosage of Drug X in 2023") were characterized by a high semantic certainty (0.82±0.12) and showed concentrated embeddings. Finder achieved 85.8\% top-10 accuracy, therefore, it could be considered quite reliable in the regulatory workflows.

    \item \textbf{Conceptual Queries} (e.g., "Challenges in oncology trial recruitment") were held at a medium level of certainty (0.71±0.18) and had coherent but less bounded embeddings. Finder's top-10 accuracy was 73.3\%, which means that it had a solid but somewhat less stable performance.

    \item \textbf{Ambiguous Queries} (e.g., "What happened in 2023?") had low certainty (0.49±0.21) with dispersed embeddings. Consequently, the accuracy of Finder decreased to 56.3\%. Intent parsing by LLM, synonym expansion, and auto-filtering helped, however, to coalesce many of the "Not Found" cases into one group.
\end{itemize}

In general, the Finder system is a dependable one for factual and conceptual queries, the main ones pharma use cases, as it shows in the tests. Ambiguous queries still hamper the system greatly, but the results are made better by the use of adaptive techniques such as dynamic reformulation and re-ranking. Semantic certainty scoring, which will be part of future versions, will enable the identification of uncertain queries at an early stage and will allow for the implementation of corrective strategies.

\begin{table}[htbp]
\caption{Semantic Certainty Benchmarking Results}
\label{tab:semantic_certainty}
\begin{center}
\begin{tabular}{|p{1.25cm}|p{1.75cm}|p{1.75cm}|p{2.25cm}|}
\hline
\textbf{Query Type} & \textbf{Avg. Certainty Score} & \textbf{Finder Accuracy (Top-10)} & \textbf{Observations} \\
\hline
Factual & $0.82 \pm 0.12$ & 0.858 & Most reliable embeddings; stable semantic wells. \\
\hline
Conceptual & $0.71 \pm 0.18$ & 0.733  & Moderate certainty; consistent but less bounded. \\
\hline
Ambiguous & $0.49 \pm 0.21$ & 0.563 & Lowest certainty; mitigated via intent parsing and auto-filtering. \\
\hline
\end{tabular}
\end{center}
\end{table}

\section{Results and Observations}

This part of the paper informs about the quantitative and qualitative performance of the \textit{Finder} platform on enterprise-scale pharmaceutical retrieval tasks. The test set comprising 1,000 queries, which spanned multimodal document types, was designed to reflect regulatory, scientific, and commercial use cases, and the evaluations were carried out on it.

\subsection{Quantitative Metrics}

Finder demonstrated competitive retrieval effectiveness on standard IR metrics. Table IV summarizes the precision, recall and ranking-based measures observed at multiple cutoff points.

\begin{table}[htbp]
\caption{Retrieval Performance Metrics}
\label{tab:retrieval_metrics}
\begin{center}
\begin{tabular}{|c|c|c|c|c|}
\hline
\textbf{Metric} & \textbf{@5} & \textbf{@10} & \textbf{@20} & \textbf{@30} \\
\hline
Precision & 0.7933 & 0.7833 & 0.7083 & 0.6833 \\
\hline
Recall    & 0.1427 & 0.2685 & 0.4493 & 0.6510 \\
\hline
nDCG      & 0.8063 & 0.8008 & 0.7670 & 0.7935 \\
\hline
\end{tabular}
\end{center}
\end{table}

Besides the usual retrieval metrics, aggregate measures also confirm Finder's effectiveness. The system got a Mean Reciprocal Rank (MRR) of 0.9014 and a Mean Average Precision (MAP) of 0.7642, pointing to its capacity of repeatedly finding very relevant documents among the top ranks. Over the entire evaluation set, Finder was able to find at least one relevant document for 87.6\% of all queries, thus showing good coverage across different modalities. Table~\ref{tab:relevant_distribution} gives an overview of the distribution of relevant documents over the different rank ranges, revealing that 82.7\% of relevant items were found in the top-10 results, with lower percentages in the subsequent ranks.

\begin{table}[htbp]
\caption{Relevant Document Distribution by Rank Range}
\label{tab:relevant_distribution}
\begin{center}
\begin{tabular}{|c|c|}
\hline
\textbf{Rank Range} & \textbf{\% of Documents} \\
\hline
Top 10  & 82.7\% \\
\hline
Top 20  & 1.9\% \\
\hline
Top 50  & 2.0\% \\
\hline
Top 100 & 1.1\% \\
\hline
\textbf{Total Relevant Documents Found} & \textbf{87.7\%} \\
\hline
\end{tabular}
\end{center}
\end{table}

These results validate that Finder consistently finds relevant documents in the top-10 positions, which makes it a suitable tool for posting searches in the real world, where users generally only look at the first page of results.

\subsection{Processing Time Breakdown}

Finder’s multimodal ingestion and retrieval pipeline maintained low-latency performance across diverse file types:

\begin{itemize}
    \item \textbf{PDFs:} \textasciitilde193 sec end-to-end (extraction, tagging, vectorization)
    \item \textbf{Audio:} \textasciitilde116 sec (transcription + tagging)
    \item \textbf{Video:} \textasciitilde203 sec (transcription + summarization + tagging)
\end{itemize}

For querying:

\begin{itemize}
    \item \textbf{Query Handling Latency:} \textasciitilde2 sec (optimization) + \textasciitilde5 sec (API response)
    \item \textbf{System Throughput:} Up to 200 concurrent queries and 32 parallel file ingestions
\end{itemize}

These benchmarks demonstrate that the pipeline is optimized for both real-time querying and scalable ingestion at enterprise scale.

\subsection{Qualitative Feedback}

Usability logs (from medical affairs, clinical documentation, and commercial teams) plus stakeholder interviews indicated the following:

\begin{itemize}
    \item \textbf{Improved Search Trust:} Trust in retrieved results was made higher with the main focus on the context and metadata-aware filtering.
    \item \textbf{Usability Gains:}  Structured query parsing and auto-filtering made it easier for non-technical users to use the product.
    \item \textbf{Multilingual Adoption:} The adoption of native multilingual content allowed teams from all over the world to use the product.
\end{itemize}

\subsection{Observed System Strengths}

\begin{itemize}
    \item \textbf{Hybrid Search Fusion:} Integration of BM42 with Mixedbread embeddings improved both precision and recall.
    \item \textbf{LLM-Based Query Understanding:} Reduced failure cases in vague or incomplete queries.
    \item \textbf{Storage Tiering:} Delivered sub-second retrieval while ensuring cost efficiency for large-scale deployments.
\end{itemize}

\section{Business and Technical Impact}

The pharmaceutical industry is a heavily regulated and data-dense environment where multimodal content that is easily accessible and quickly interpretable forms the basis of sound decision-making, compliance processes, and operational agility. Unfortunately, traditional search systems have been notoriously ineffective at meeting the growing demand for holistic data access, as they typically lack the necessary support for non-textual data. This situation leads to assets going untapped and a manual effort that increases over time.

\textit{Finder}  solves the problems mentioned above by going beyond the traditional file system to provide a one-stop access to all kinds of documents, images, audio, and video through AI-powered ingestion, enrichment, and hybrid retrieval. By doing so, the system not only improves findability but also eliminates the re-use of files and thus saves time and money in the organization’s entire regulatory, clinical, and commercial activities.

\subsection{Enterprise Value Realization}

Initial deployments of Finder demonstrated measurable impact:

\begin{itemize}
    \item \textbf{Content Search and Retrieval:} \textasciitilde40\% reduction in time to locate documents; \textasciitilde35\% increase in search increase in relevance.
    \item \textbf{Content Reusability:} \textasciitilde30\% less effort in retrieving reusable content for brand planning.
    \item \textbf{Compliance Support:} Automated tagging saved \textasciitilde50 hours/month in metadata curation.
    \item \textbf{Workflow Efficiency:} Manual inspection effort reduced by up to \textasciitilde45\%.
\end{itemize}

\subsection{Operational Scalability}

Finder’s architecture is built for horizontal scaling and enterprise-grade reliability:

\begin{itemize}
    \item \textbf{Parallelized Processing:} Supports ingestion of large volumes across distributed workers.
    \item \textbf{Tiered Storage:} Optimized via Qdrant for balancing performance and cost.
    \item \textbf{Auto-Healing Pipelines:} Automatically recover from ingestion or parsing failures with retry logic and alerting.
\end{itemize}

During enterprise user acceptance testing (UAT), Finder sustained up to 200 concurrent queries and 32 concurrent file ingestions, with sub-2 second latency for 95\% of requests.

\section{Conclusion}

This paper presents \textit{Finder}, a scalable and multimodal AI-powered search framework designed for enterprise-grade pharmaceutical content discovery. The described work introduces \textit{Finder}, a Forge-powered, scalable and multimodal AI-based search framework for the discovery of pharmaceutical-content with enterprise-grade. Not only does Finder meld traditional keyword-based retrieval with LLM (large language model)-powered semantic reasoning, but it also facilitates a unified access to the various and heavily-regulated data formats such as documents, audio, video, and images from clinical, regulatory and commercial domains.

The concept of the system is the synergy of a modular pipeline for ingestion, a hybrid architecture for the retrieval of data, and a metadata-aware ranking strategy for providing the users with accurate search results and at the same time keeping the scalability of the operations. The melding of BM42 lexical scoring with Mixedbread semantic embeddings, complemented by fuzzy matching and LLM-based query optimization, is what maintains the top ranking of the found documents even when the query is complicated and domain-specific.

The qualitative study on synthetic benchmarks and real-world enterprise queries proves the reliability of Finder, scoring an average of 87.7\% of the relevance rate across 1,000 test queries. Its implementation has resulted in several positive changes in content discoverability, reuse, and compliance readiness, as well as an easy support for high-throughput ingestion and low-latency retrieval in the production environment.

In the future, Finder will remain as a solid base for new developments in the corporate search field. Upgrades to come will mainly concentrate on personalized ranking, adaptive feedback loops, as well as more profound connections with decision-support systems, thus widening its reach beyond pharmaceutical research, regulatory intelligence, and knowledge management.

\section*{Acknowledgment}

We sincerely thank Samik Adhikary, Puneet Srivastava, Jennifer McGuire, Saurabh Oberoi, and Denis Mayer for their invaluable support and subject matter expertise.


\begin{thebibliography}{00}

\bibitem{b1} K. Adnan, R. Akbar, S. W. Khor, and A. B. A. Ali, ``Role and challenges of unstructured big data in healthcare,'' *Data Management, Analytics and Innovation*, pp. 301–323, 2020. [Online]. 

\bibitem{b2} S. Olsson, S. N. Pal, and A. Dodoo, ``Pharmacovigilance in resource-limited countries,'' *Expert Rev. Clin. Pharmacol.*, vol. 8, no. 4, pp. 449–460, 2015. [Online]. 

\bibitem{b3} Y. Wang et al., ``Exploring the reasoning abilities of multimodal large language models (MLLMs): A comprehensive survey on emerging trends in multimodal reasoning,'' *arXiv preprint arXiv:2401.06805*, 2024. [Online]. 

\bibitem{b4} OpenAI, ``Introducing ChatGPT: Optimizing Language Models for Dialogue,'' *OpenAI Blog*, Nov. 30, 2022.

\bibitem{b5} A. Radford et al., ``Robust Speech Recognition via Large-Scale Weak Supervision,'' OpenAI, 2022.

\bibitem{b6} TwelveLabs, ``Platform Overview – Video Understanding Platform,'' *TwelveLabs Documentation*, 2023.

\bibitem{b7} S. Agarwal et al., ``Evaluating CLIP: Towards Characterization of Broader Capabilities and Downstream Implications,'' *arXiv preprint arXiv:2108.02818*, 2021.

\bibitem{b8} J. Li, D. Li, C. Xiong, and S. Hoi, ``BLIP: Bootstrapping Language-Image Pre-training for Unified Vision-Language Understanding and Generation,'' in *Proc. ICML*, pp. 12888–12900, 2022.

\bibitem{b9} J. B. Alayrac et al., ``Flamingo: A Visual Language Model for Few-Shot Learning,'' *Advances in Neural Information Processing Systems*, vol. 35, pp. 23716–23736, 2022.

\bibitem{b10} A. Hurst et al., ``GPT-4o System Card,'' *arXiv preprint arXiv:2410.21276*, Oct. 2024.

\bibitem{b11} Gemini Team et al., ``Gemini: A Family of Highly Capable Multimodal Models,'' *arXiv preprint arXiv:2312.11805*, 2023.

\bibitem{b12} S. E. Robertson and S. Walker, ``Some Simple Effective Approximations to the 2-Poisson Model for Probabilistic Weighted Retrieval,'' in *Proc. SIGIR*, pp. 232–241, 1994.

\bibitem{b13} V. Karpukhin et al., ``Dense Passage Retrieval for Open-Domain Question Answering,'' in *Proc. EMNLP*, pp. 6769–6781, 2020.

\bibitem{b14} T. Formal, B. Piwowarski, and S. Clinchant, ``SPLADE: Sparse Lexical and Expansion Model for First Stage Ranking,'' in *Proc. 44th ACM SIGIR*, pp. 2288–2292, 2021.

\bibitem{b15} W. C. Chang et al., ``Pre-training Tasks for Embedding-Based Large-Scale Retrieval,'' *arXiv preprint arXiv:2002.03932*, 2020.

\bibitem{b16} H. K. Azad and A. Deepak, ``Query Expansion Techniques for Information Retrieval: A Survey,'' *Information Processing \& Management*, vol. 56, no. 5, pp. 1698–1735, 2019.

\bibitem{b17} J. Kim, M. Hur, and M. Min, ``From RAG to QA-RAG: Integrating Generative AI for Pharmaceutical Regulatory Compliance Process,'' in *Proc. 40th ACM/SIGAPP SAC*, pp. 1293–1295, 2025.

\bibitem{b18} Y. Yun et al., ``EICopilot: Search and Explore Enterprise Information over Large-scale Knowledge Graphs with LLM-driven Agents,'' *arXiv preprint arXiv:2501.13746*, 2025.

\bibitem{b19} F. Holzschuher and R. Peinl, ``Performance of Graph Query Languages: Comparison of Cypher, Gremlin and Native Access in Neo4j,'' in *Proc. Joint EDBT/ICDT Workshops*, pp. 195–204, 2013.

\bibitem{b20} M. Douze et al., ``The FAISS Library,'' *arXiv preprint arXiv:2401.08281*, 2024.

\bibitem{b21} S. Bruch et al., ``Bridging Dense and Sparse Maximum Inner Product Search,'' *ACM Trans. Inf. Syst.*, vol. 42, no. 6, pp. 1–38, 2024.

\bibitem{b22} J. Wang et al., ``Milvus: A Purpose-Built Vector Data Management System,'' in *Proc. SIGMOD*, pp. 2614–2627, 2021.

\bibitem{b23} Y. A. Malkov and D. A. Yashunin, ``Efficient and Robust Approximate Nearest Neighbor Search Using Hierarchical Navigable Small World Graphs,'' *IEEE Trans. Pattern Anal. Mach. Intell.*, vol. 42, no. 4, pp. 824–836, 2018.

\bibitem{b24} Amazon, ``Amazon Nova Lite Foundation Model (amazon.nova-lite-v1:0),'' *AWS Bedrock Documentation*, 2024.

\bibitem{b25} RapidFuzz Documentation, ``rapidfuzz.fuzz.token\_set\_ratio,'' *RapidFuzz v3.13.0*, 2025. [Online]. Available: RapidFuzz fuzz usage documentation.


\bibitem{b26} mixedbread-ai, ``Mixbread model: mxbai-embed-large-v1,'' *Hugging Face*, [Online]. Available: https://huggingface.co/mixedbread-ai/mxbai-embed-large-v1

\bibitem{b27} Qdrant, ``Bm25,'' *Hugging Face*, [Online]. Available: https://huggingface.co/Qdrant/bm25

\bibitem{b28} Qdrant, ``Bm42:all\_miniLM\_L6\_v2\_with\_attentions,'' *Hugging Face*, [Online]. Available: https://huggingface.co/Qdrant/all\_miniLM\_L6\_v2\_with\_attentions

\bibitem{b29} Y. Du, ``Semantic Certainty Assessment in Vector Retrieval Systems: A Novel Framework for Embedding Quality Evaluation,'' *arXiv preprint arXiv:2507.05933*, 2025. [Online]. Available: https://arxiv.org/abs/2507.05933

\end{thebibliography}
\end{document}